\documentclass[12pt,a4paper]{article}
\usepackage{lmodern}
\usepackage[T1]{fontenc}
\usepackage[utf8]{inputenc}

\title{\bfseries Geometric Phases for Mixed States\\of the Kitaev Chain}

\author{\normalsize Ole Andersson\thanks{\texttt{ole.andersson@fysik.su.se}}\,\, and Ingemar Bengtsson\thanks{\texttt{ingemar.bengtsson@fysik.su.se}} \\
\small\em Department of Physics, Stockholm University, \\
\small\em SE-106 91 Stockholm, Sweden\vspace{5pt}\\
\normalsize Marie Ericsson\thanks{\texttt{marie.ericsson@physics.uu.se}} \,\, and Erik Sj\"oqvist\thanks{\texttt{erik.sjoqvist@physics.uu.se}}\\
\small\em Department of Quantum Chemistry, Uppsala University,\\
\small\em Box 518, SE-751 20 Uppsala, Sweden\vspace{3pt}\\
\small\em Department of Physics and Astronomy, Uppsala University,\\
\small\em Box 516, SE-751 20 Uppsala, Sweden}

\date{\normalsize\today}

\usepackage{color, amsmath, amsthm, amsfonts, bbm}
\usepackage[pdftex]{graphicx}
\usepackage{epstopdf}
\usepackage{subfig}

\newcommand{\HH}{\mathcal{H}}
\newcommand{\A}{\mathcal{A}}
\newcommand{\Tr}{\operatorname{Tr}}
\renewcommand{\phi}{\varphi}

\newcommand{\ket}[1]{|{#1}\rangle}

\newcommand{\braket}[2]{\langle #1|#2\rangle}
\newcommand{\ketbra}[2]{|#1\rangle\langle #2|}
\DeclareMathOperator{\sech}{sech}

\begin{document}
\maketitle
\vspace{-20pt}
\begin{abstract}
\noindent The Berry phase has found applications in building topological 
order para\-meters for certain condensed matter systems. The question 
whether some geometric phase for mixed states can serve the same purpose 
has been raised, and proposals are on the table. We analyze the intricate 
behaviour of Uhlmann's geometric phase in the Kitaev chain at finite temperature, and then argue that
it captures quite different physics from 
that intended. We also analyze the behaviour of a geometric phase introduced in the context of interferometry. For the Kitaev chain, this phase closely mirrors that of the Berry phase, and we argue that it merits 
further investigation. 
\end{abstract}

\newpage

\section{\large Introduction}
Quantum theory offers a rich supply of mathematical concepts, and, in principle, each of these raises the question of its physical interpretation. The geometric phase is an interesting example of such a concept as new applications of various geometric phases continue to appear.
We have discussed this circle of ideas before, in rather general terms \cite{Erikvaxjo2002,Ingemarvaxjo2005}. Here 
our intention is to see how they play out in a context with specific demands on the interpretation. 

The Berry phase for pure states has found application in the study of topologically 
ordered matter, and recently there has been some discussion \cite{Viyuela_etal2014a,Viyuela_etal2014b,Viyuela_etal2014c,Huang_etal2014,Budich_etal2015}
about the use of geometric phases for mixed states, and the role they can play, when topologically ordered matter is kept at finite temperature.  
In this paper we analyze the behaviour of 
two geometric phases for mixed states---Uhlmann's phase \cite{Uhlmann1986,Uhlmann1989,Uhlmann_etal2009} and the inequivalent \cite{Slater2001} interferometric phase \cite{Sjoqvist_etal2000,Tong_etal2004}---in the Kitaev chain, which is a simple model that can undergo a topological phase transition. For the Uhlmann phase, we find a behaviour that is surprisingly intricate. However, we will eventually argue that the behaviour is not really tied to the physics we want to capture.
In contrast to the Uhlmann phase, the interferometric geometric phase 
adopts a topological character and makes a discrete jump when the multiplicity of the   
spectrum of the state changes. In our calculation, this translates into a 
jump when the band gap closes in the Kitaev chain. Thus it is potentially a useful topological invariant, and we will argue that it deserves further study in the context.

The paper is organized as follows. In Section \ref{Kitaev}, we introduce the Kitaev chain
and its associated Gibbs states, and in Sections \ref{UhlmannGP} and \ref{betong} we define and analyze Uhlmann's geometric phase for these. Section \ref{interGP} contains a contrasting analysis of the interferometric phase, and Section \ref{discussion} contains a discussion and our conclusions.

\section{\large The Kitaev chain}\label{Kitaev}
A brief introduction to topological phases of matter may begin by reminding 
the reader that phases of matter have been very successfully understood by 
means of the concept of a local order parameter, as in Landau's theory. The 
new twist is that in certain materials the order parameter may take on a global, 
topological character. Indeed, the issue here is whether some particular geometric phase can serve as an order parameter for a topological phase.\footnote{The 
use of the word `phase' in two different senses is unavoidable, and can be traced back to Gibbs.} 

Let us consider a simple model known as the Kitaev chain 
\cite{Kitaev2001} in which such behaviour can be seen. It is a $1$-dimensional 
model built from spinless fermions ($a_na_m^\dagger + a_m^\dagger a_n = 
\delta_{nm}$) at $N$ sites with the Hamiltonian 
\begin{equation}\label{kitaev hamiltonian} 
H = \sum_{n=1}^{N}\left[ - w(a_n^\dagger a_{n+1} + a_{n+1}^\dagger a_n) - \mu 
a_n^\dagger a_n + M (a_na_{n+1} + a_{n+1}^\dagger a_n^\dagger ) \right]. 
\end{equation} 
Here $w$ is the hopping amplitude, $\mu$ is the chemical potential and 
$M$ is known as the induced superconducting gap. 
We assume that the number of sites is large and that periodic boundary 
conditions are used. 

We can use a Fourier transformation 
to reexpress the Hamiltonian as 
\begin{equation}
H = \int_0^{2\pi}\frac{\operatorname{d}\!k}{2\pi} \Psi_k^\dagger H_k \Psi_k,
\end{equation}
where the Nambu spinor is $\Psi_k = (a_k , a_{-k}^\dagger )^{\rm T}$ and 
\begin{equation} 
H_k = - M \sin{k}\sigma_y - \left( \frac{\mu}{2} + w\cos{k}\right)\sigma_z.
\end{equation}
The eigenvalues of this Hamiltonian are 
\begin{equation} 
\lambda_\pm = \pm \sqrt{\left( \frac{\mu}{2} + w\cos{k}\right)^2 + 
\left( M \sin{k}\right)^2}. 
\end{equation}
Here, to simplify matters, we will set $w=M=1$ and $m=\mu/2$.
Moreover, for convenience, we will represent the Hamiltonian so that the Bloch vector lies in the equatorial plane. Thus, we consider Hamiltonians of the form 
\begin{equation}
H_k = - \frac{\Delta_k}{2}\vec{n}_k\cdot \vec{\sigma}, 
\end{equation} 
where $\vec{\sigma} = (\sigma_x, \sigma_y, \sigma_z)$ are the Pauli matrices and 
\begin{align}
\vec{n}_k &= \frac{2}{\Delta_k}\left( m+\cos{k},\sin{k},0\right),\label{ily}\\ 
\Delta_k &= 2\sqrt{(m+\cos{k})^2 + \sin^2{k}}. \label{fam}
\end{align}

\begin{figure}[t]
\centering
\includegraphics[width=0.80\textwidth]{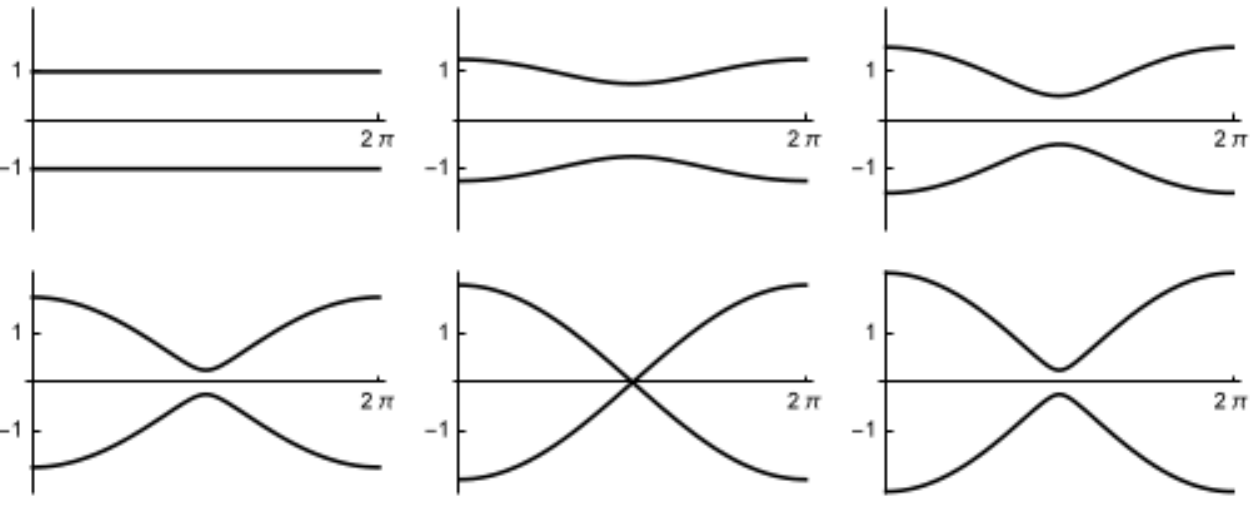}
\caption{\small The spectra for $\Delta_k$ as in Eq.~\eqref{fam} 
and $m = 0, 1/4, 2/4$ from left-to-right in the top row and $m=3/4, 1, 5/4$ from left-to-right in the bottom row. The fact that it matters whether $m$ is larger 
or smaller than one is not yet apparent.} 
\label{fig:illo7}
\end{figure}

What is the physics of this? The model has two energy bands (which is why we can treat 
it as a qubit), with a band gap that closes if $m = 1$. See Fig.~\ref{fig:illo7}. Kitaev 
found that in an open chain with a finite number of sites, the model behaves very differently 
depending on whether $m < 1$ or $m>1$. In the former case, there are two extra states within 
the gap, which turn out to be exponentially localized on the edges of the chain. In 
the latter case, these states are missing. In this sense, there is a kind of phase transition at $m = 1$. 

\begin{figure}[ht]
\centering
\includegraphics[width=0.80\textwidth]{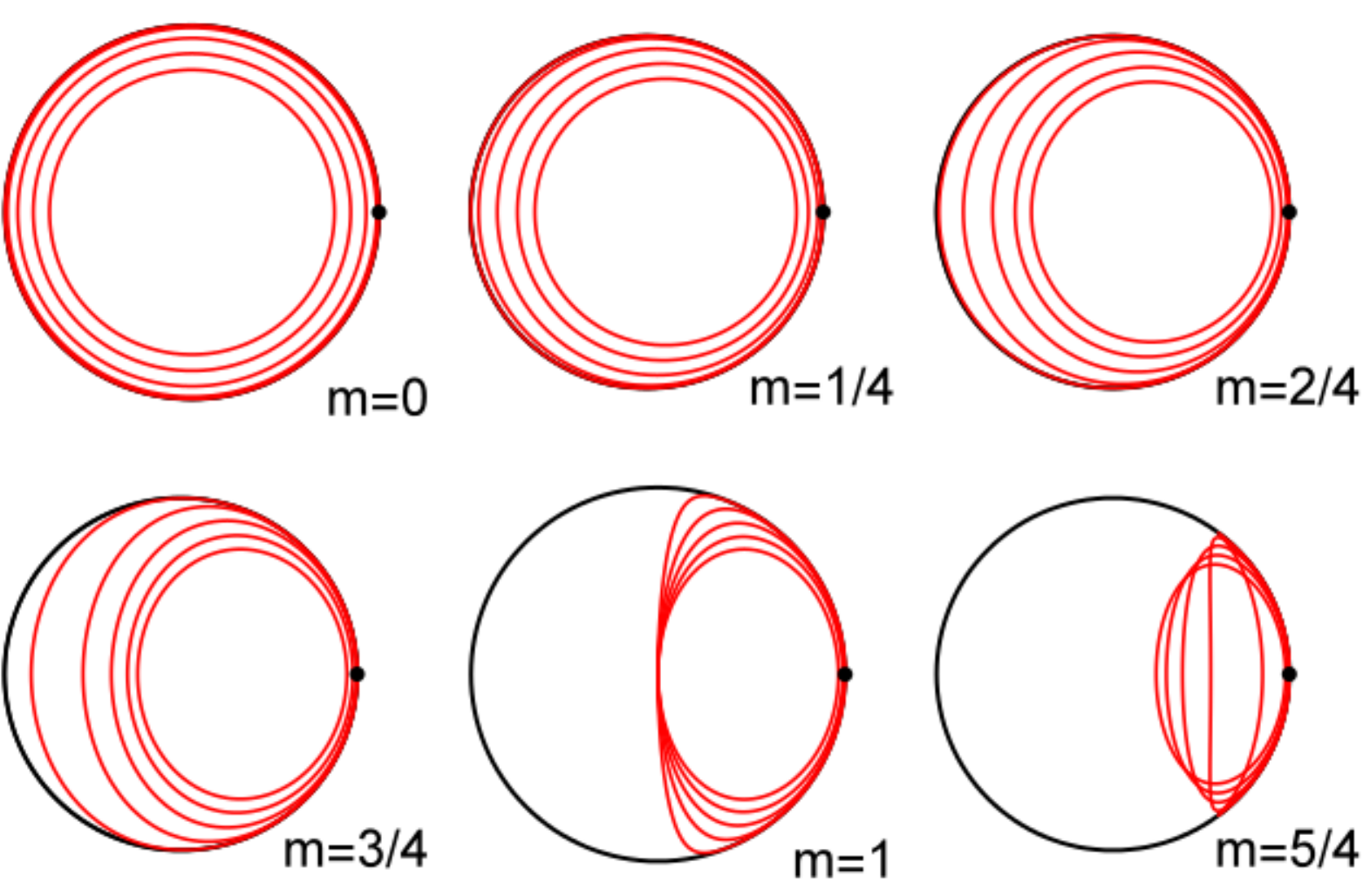}
\caption{\small Curves in the Bloch ball for some values of $m$. 
In each panel, we see the curves swept out by the Gibbs states as we move along the Brillouin zone 
for $T = 1/5, 2/5, 3/5, 4/5, 1$. The east pole is marked with a dot.} 
\label{fig:illo3}
\end{figure}

In our model, we can discern a topological order parameter by looking 
at the Bloch vector \eqref{ily}. As we move along the Brillouin 
zone, parameterized by $k$, the eigenstates move around the equator of the 
Bloch sphere for $m<1$, whereas for $m>1$ they are rocking back and forth around the east 
pole. In the former case, we pick up a Berry phase factor 
$e^{i\pi}$; in the latter case, the Berry phase is zero \cite{Berry1984,Aharonov_etal1987,Zak1989}. For $m=1$, the gap closes and 
the Berry phase is ill-defined. Thus, we can say that there is a 
topological $\mathbb{Z}_2$ invariant taking different values in the two phases, and this invariant 
serves as the order parameter of the model. A systematic classification of non-interacting models of this kind is available \cite{Schnyder_etal2008,Qi_etal2008,Kitaev2009,Ryu_etal2010}.

After this (absurdly brief) introduction to topological phases of matter, we can go on to 
ask about mixed states. Assume that the chain is in contact with a thermal bath at temperature $T$, and is 
described by the Gibbs state 
\begin{equation} 
\rho (k) = \frac{e^{-H_k/T}}{\mbox{Tr}e^{-H_k/T}} = 
\frac{1}{2}\left( \mathbbm{1} + \tanh{\frac{\Delta_k}{2T}}  
\vec{n}_k\cdot \vec{\sigma}\right). 
\end{equation}
This state defines a curve within the equatorial plane 
of the Bloch ball, as 
shown in Fig.~\ref{fig:illo3} for various values of $m$ and $T$. These curves can be 
described analytically. Using polar coordinates on the equatorial plane we have for the 
angular coordinate that 
\begin{equation}
\phi (k) = \frac{k}{2} + \arctan{\left( \frac{1-m}{1+m}\tan{\frac{k}{2}}\right)}.
\end{equation}
The case $m = 0$, which is called the flat band case, is particularly simple 
because the Brillouin zone is mapped to circles centred at the maximally mixed state. 
For $m<1$ the curves enclose, for $m=1$ the curves pass through and for $m>1$ the curves are confined to one side of the maximally mixed state.

The question now is if we can define a phase factor for mixed states, analogous to the 
Berry phase for pure states, which can serve as a topological invariant describing the 
phase structure of the Kitaev chain at finite temperature. The work by Viyuela \emph{et al.}~\cite{Viyuela_etal2014a, Viyuela_etal2014c} at first sight 
suggests that the Uhlmann geometric phase \cite{Uhlmann1986} can play this role. It would further  
suggest that there is a sharp transition between different phases of the model also at 
finite $T$, and that this transition happens at an $m = m_c(T) < 1$. For 
curves in the equatorial plane of the Bloch ball the Uhlmann phase factor is indeed 
a number $\pm 1$ which can be calculated, and by fixing the curves in this manner 
Viyuela \emph{et al.}~have added an intriguing amount of concreteness to its study. We will add 
some more, and discuss its interpretation afterwards. We will also broaden the perspective 
slightly and study the behaviour of another candidate geometric phase for mixed states 
\cite{Sjoqvist_etal2000}. A paper with a similar aim has appeared recently \cite{Budich_etal2015}.

\section{\large The Uhlmann phase factor}\label{UhlmannGP}
The basic idea in Uhlmann's theory is to let the pure states in the bipartite Hilbert space $\HH\otimes\HH^\star$
form the total space of a fiber bundle over the mixed states on $\HH$, see Ref.~\cite{Uhlmann_etal2009}.
The projection onto the mixed states is $\psi \rightarrow \psi \psi^\dagger = \rho$, and the structure 
group is $U(N)$ acting from the right, $\psi\rightarrow \psi U$.
A geometric phase can be associated to any curve in the base manifold once we have defined a parallelism condition for curves in the total space---a lift $\psi(k)$ of $\rho(k)$ is said to be parallel if for every infinitesimal $\delta k$ the probability for the transition from $\psi(k)$ to $\psi(k+\delta k)$  
equals the fidelity of $\rho(k)$ and $\rho(k+\delta k)$, 
\begin{equation}
\left|\Tr(\psi(k)^\dagger \psi(k+\delta k))\right|^2=\Tr\sqrt{\rho (k)^{1/2}\rho(k+\delta k)\rho (k)^{1/2}}.
\end{equation}

The parallelism condition can also be described in terms of a connection \cite{Uhlmann1991}, which we here denote by $\A$. 
In general, no explicit formula for $\A$ is known. However, H\"ubner \cite{Hubner1993} has derived a formula giving its values along the velocity fields of 'square root lifts' $\psi=\sqrt{\rho}$.
The formula of H\"ubner reads 
\begin{equation} 
\mathcal{A}(\dot{\psi}) = \sum_{i,j}|u_i\rangle \frac{\langle u_i| 
[\dot{\psi},\psi]|u_j\rangle }{p_i+p_j}\langle u_j|,\label{hubner} 
\end{equation}
where the $p_i$ and the $\ket{u_i}$ are the eigenvalues and eigenstates of $\rho$, and the overdot means differentiation with respect to $k$.
It should be pointed out that the derivation of \eqref{hubner} assumes that $\rho$ has full rank, so it fails when the system is in a 
pure state. Indeed, the Uhlmann bundle is defined for mixed states 
of full rank only. Apart from that, it is insensitive to the spectra of the mixed states. 

Let us postpone a discussion of the physical import of Uhlmann's construction and turn directly 
to an evaluation of the Uhlmann phase for curves in the equatorial plane of the Bloch ball. 
In general, H\"ubner's formula \eqref{hubner} is hard to handle, but for the qubit 
state space it simplifies to 
\begin{equation} 
\mathcal{A}(\dot{\psi}) = (\sqrt{p_1}-\sqrt{p_2})^2(|u_1\rangle 
\langle u_1|\dot{u}_2\rangle \langle u_2| + |u_2\rangle \langle u_2|\dot{u}_1\rangle 
\langle u_1|).
\end{equation}
Since $\rho$ depends on $k$, so do its eigenvectors $\ket{u_i}$ and its 
eigenvalues $p_i$. We are interested in curves that stay in the equatorial plane of the Bloch 
ball, and set 
\begin{equation}\label{mixat}
\rho = \frac{1}{2}\begin{pmatrix}1 & (p_1-p_2)e^{-i\phi} \\ (p_1-p_2)e^{i\phi} & 1 \end{pmatrix}
\end{equation}
and 
\begin{equation}
\ket{u_1} = \frac{1}{\sqrt{2}}\begin{pmatrix} 1 \\ e^{i\phi} \end{pmatrix},\qquad \label{ren1}
\ket{u_2} = \frac{1}{\sqrt{2}}\begin{pmatrix} 1 \\ -e^{i\phi} \end{pmatrix}. 
\end{equation}
Then the connection becomes abelian, 
\begin{equation} 
\mathcal{A}(\dot{\psi}) = \frac{i}{2}\dot{\phi} 
(\sqrt{p_1}-\sqrt{p_2})^2
\begin{pmatrix}
-1 & 0 \\ 0 & 1
\end{pmatrix},
\end{equation}
and no path ordering is necessary in order to compute the symmetry group element 
\begin{equation} 
U(k) = \exp\left( -\int_0^k\!\operatorname{d}\!k\, \mathcal{A}(\dot{\psi})\right) 
= \begin{pmatrix}
e^{iA} & 0 \\ 0 & e^{-iA}
\end{pmatrix}, 
\end{equation}
where 
\begin{equation} 
A = \frac{1}{2}\int_0^k\!\operatorname{d}\!k\,\dot{\phi}(\sqrt{p_1}-\sqrt{p_2})^2. \label{A1} 
\end{equation}
A parallel lift of $\rho(k)$, then, is
\begin{equation} 
\psi_{||}(k) 
= \psi (k)U(k) 
= \frac{1}{2}\begin{pmatrix}
(\sqrt{p_1} + \sqrt{p_2})e^{iA} & (\sqrt{p_1}-\sqrt{p_2})e^{i(A-\phi )} \\
(\sqrt{p_1} - \sqrt{p_2})e^{i(\phi - A)} & (\sqrt{p_1}+\sqrt{p_2})e^{-iA} 
\end{pmatrix}. \label{para} 
\end{equation}
The unitary $U(k)$ depends on the special lift we started out with, $\psi=\sqrt{\rho}$,
but $\psi_{||}(k)$ does not. Let us remark---since it will be important to us later on---that 
$\psi_{||}(k)$ is in general not periodic, even if the original curve 
$\rho (k)$ is closed with periodicity $2\pi$. The function $A(k)$ need not be periodic 
even if $\phi (k)$ is. 

Uhlmann's geometric phase is the argument of the phase factor of the function 
\begin{equation}
\begin{split}
\Tr\left(\psi_{||}(0)^\dagger \psi_{||}(k)\right)&
= \frac{1}{2}\Big(\sqrt{p_1(0)}+\sqrt{p_2(0)}\Big)\Big(\sqrt{p_1(k)}+\sqrt{p_2(k)}\Big)\cos{A}\,+ 
\\  &+ \frac{1}{2}\Big(\sqrt{p_1(0)}-\sqrt{p_2(0)}\Big)\Big(\sqrt{p_1(k)}-\sqrt{p_2(k)}\Big)
\cos{(\phi+A)}. 
\end{split}\label{holon}
\end{equation}
Because we consider curves in the equatorial plane of the Bloch ball, this is a 
real-valued quantity. For small values of $k$ therefore the phase factor equals $+1$, and 
it stays that way until we reach a node on the curve, that is, a point where the trace 
vanishes. At a non-degenerate node the phase factor jumps to $-1$. If the curve was deformed slightly away from the 
equatorial plane we would see the phase building up continuously around the position 
of the node.  

\section{\large Concrete behaviour of the Uhlmann phase factor}\label{betong}
We now specialize to the case of Gibbs states of the Kitaev chain, for which 
\begin{equation} 
p_1 = \frac{1}{2}\big(1+\tanh{\frac{\Delta_k}{2T}}\big),\qquad 
p_2 = \frac{1}{2}\big(1-\tanh{\frac{\Delta_k}{2T}}\big).
\end{equation}
To simplify matters we consider one of two cases, either the flat band 
case ($m=0$, $\Delta_k = 2$) or the case of closed curves (using $\Delta_0 = \Delta_{2\pi}$). 
In either case $p_i(k) = p_i(0) = p_i$, and Eq.~\eqref{holon} simplifies. We find that 
\begin{equation}
(\sqrt{p_1}+\sqrt{p_2})^2 = 1 + \sech\frac{\Delta_k}{2T},\qquad 
(\sqrt{p_1}-\sqrt{p_2})^2 = 1 - \sech\frac{\Delta_k}{2T}.
\end{equation}
Before inserting this result into Eq.~\eqref{holon}, we define 
\begin{equation} 
x = \sech\frac{\Delta_k}{2T}.
\end{equation}
Recalling that Uhlmann's phase factor is multiplied by a factor 
$-1$ each time the curve encounters a node, we see that the task is to 
find those values of $k$ for which 
\begin{equation}
\Tr\left(\psi (0)^\dagger \psi (k)U(k)\right) = 
\frac{1}{2}\left( (1+x)\cos{A} + 
(1-x)\cos{(\phi+A)}\right) = 0.\label{noder} 
\end{equation}
In this equation, $x = x(k;T)$ as $x$ depends on $\Delta_k$, and moreover 
\begin{equation} 
A = A(k; T) = \frac{1}{2}\int_0^k\!\operatorname{d}\!k\,\dot{\phi}(1-x). \label{A} 
\end{equation}
In general this integral is too hard for us to do analytically.

We first consider the simplest case, the flat band case in which $\Delta_k = 2$, 
$x=\sech(1/T)$ and 
\begin{equation} 
A = \frac{\phi}{2}(1-x). 
\end{equation} 
The curves are then circles, and the temperature alone determines 
the purity of the states. The limit $T\to 0$ can now be taken. When $T = 0$ the 
hyperbolic secant  vanishes, the state is pure, $A = \phi /2$ and the nodes are determined by the equation
\begin{equation} 
\cos{\frac{\phi}{2}} + \cos{\frac{3\phi}{2}} = 0
\quad\Leftrightarrow\quad
\cos{\frac{\phi}{2}}\left( \cos^2{\frac{\phi}{2}} - 
\frac{1}{2}\right) = 0. 
\end{equation}
Hence there are nodes at $\phi = \pi/2, \pi, 3\pi/2$. This should be compared with the 
the Berry phase, which has a single node at $\phi = \pi$. However, 
for a closed circle the contributions from the two extra nodes will cancel each other, 
and we obtain the same value for the Uhlmann and Berry phases (as noted by Uhlmann \cite{Uhlmann1986}).

\begin{figure}[t]
	\centering
	\subfloat[At most one turn.\label{fig:illo1a}]{%
		\includegraphics[width=0.35\textwidth]{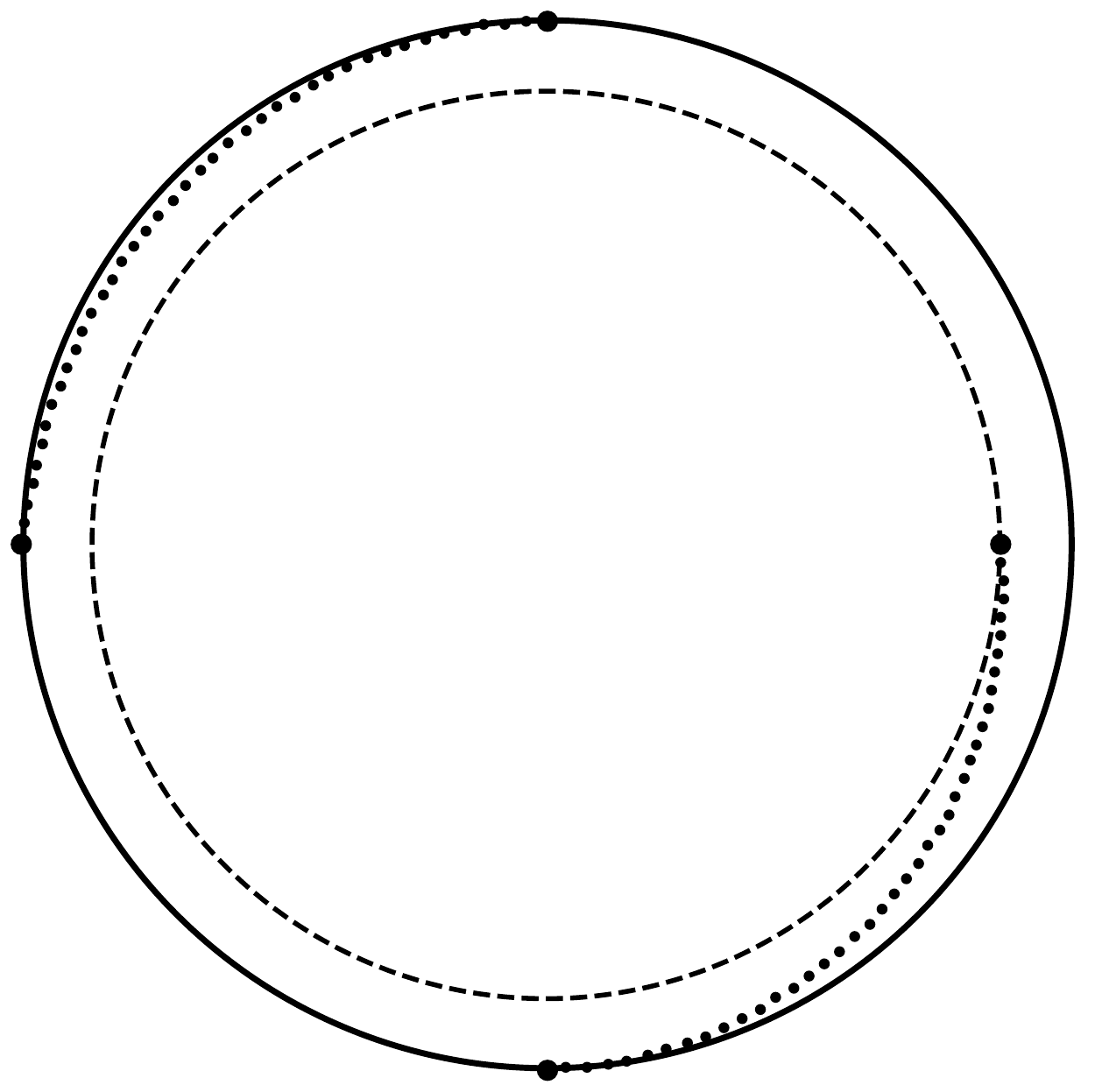}
	}
	~\hspace{20pt}
	\subfloat[Between one and two turns.\label{fig:illo1b}]{%
		\includegraphics[width=0.35\textwidth]{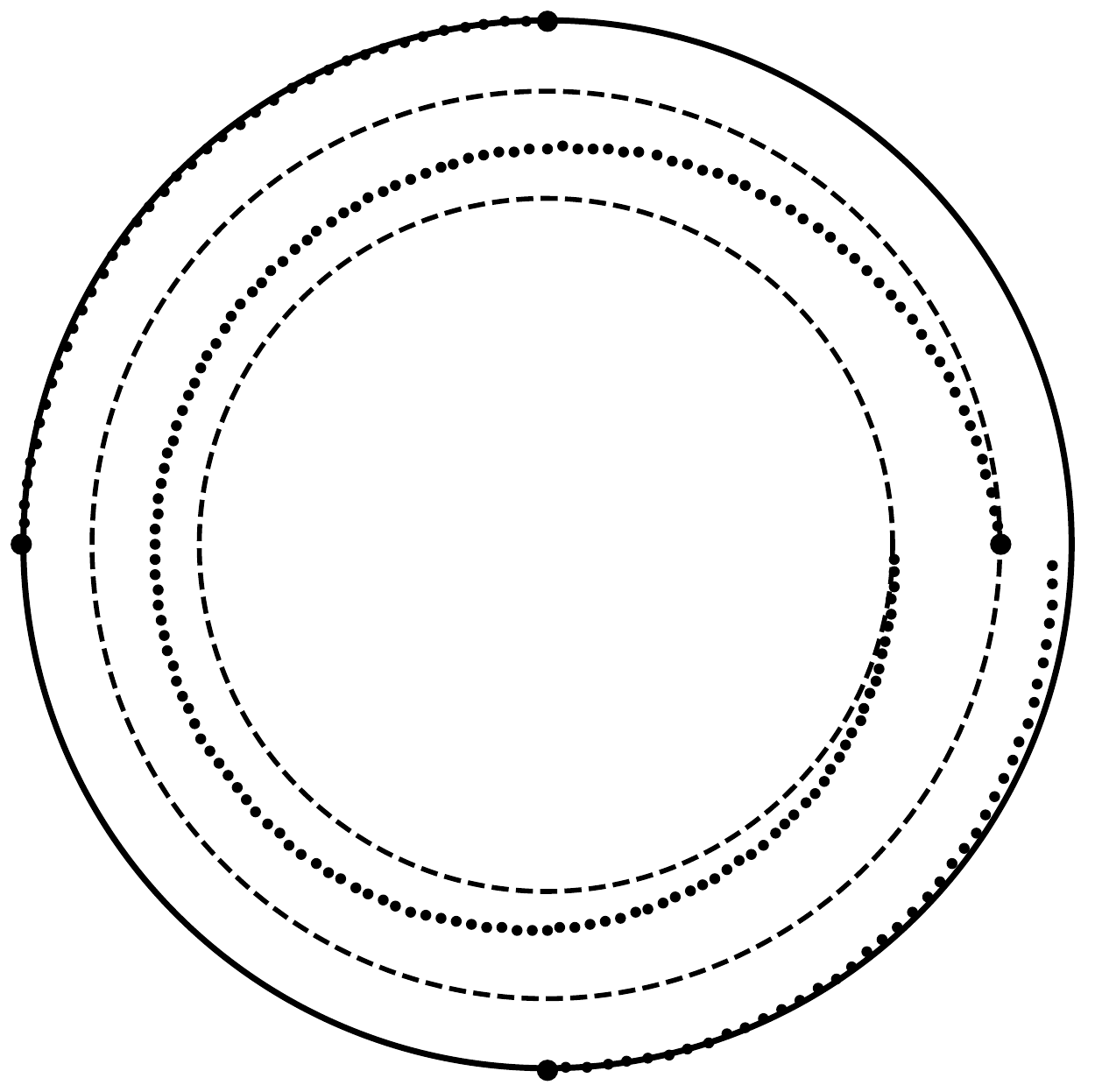}
	}
\caption{\small Where the nodes lie if the paths start on the positive $x$-axis and go 
	counter\-clockwise along circles of constant purity.} 
	\label{fig:illo1}
\end{figure}

To see what happens as we increase $T$, that is to say as we shrink the radii of 
the circles, refer to Fig.~\ref{fig:illo1a}.
The first 
thing that happens is that the two 'extra' nodes between $\pi/2$ and $\pi$
disappear. Then there is a 
value of $T$ (called the 'critical temperature' by Viyuela \emph{et al.}~\cite{Viyuela_etal2014a}) 
for which the circle coincides with the dashed circle. For higher values of $T$ no 
nodes are encountered during the first revolution around the Brillouin zone, and 
consequently Uhlmann's phase factor remains equal to $+1$. But the evolution of the 
purified state $\psi_{||}(k)$ is not periodic in $k$, so a node will be encountered 
during the second revolution (unless the temperature is so high that we are inside 
the smaller dashed circle), see Fig.~\ref{fig:illo1b}, and so on. 

\begin{figure}[t]
\centering
\includegraphics[width=0.80\textwidth]{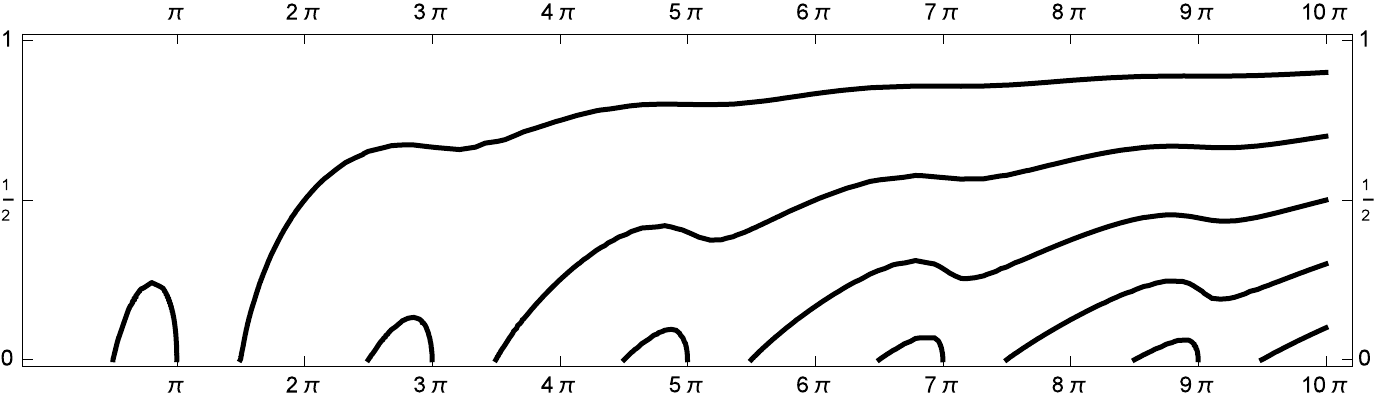}
\caption{\small Solutions of Eq.~(\ref{noder}) for $\Delta_k = 2$ and winding 
numbers up to 5, with $x =$ sech$(1/T)$ as the ordinate. For large $T$, 
that is for high entanglement, 
the circles traversed in the Bloch ball are small and several revolutions are needed to see a node.} 
\label{fig:illo4}
\end{figure}

Fig.~\ref{fig:illo4}
gives the location of the nodes for as many as 5 revolutions. 
The curves defining the nodes reflect the rather complicated evolution of $\psi_{||}(k)$. 
However, by inspection we see that some periodicities appear if we consider closed 
curves. The condition for a node to occur after exactly $n_1$ revolutions is 
\begin{equation} 
\cos\left(A(2\pi n_1)\right) = \cos((1-x)n_1\pi) = 0.
\end{equation}
There are $n_1$ solutions for $n_1$ revolutions, namely 
\begin{equation}
x_{n_1,n_2} = \frac{2(n_1-n_2)-1}{2n_1},\qquad n_2 \in \{0, 1, \dots , n_1-1\}. 
\end{equation}
The node at $x = 1/2$ repeats after two more revolutions since 
\begin{equation} 
x_{1,0} = x_{3,1} = x_{5,2} = \dots = \frac{1}{2}. 
\end{equation}  
For open curves no special simplifications are visible. 

\begin{figure}[t]
	\centering
	\subfloat[The location of the 'critical temperature' $T_{cr}=T_{1,0}$ after one turn, in agreement with \cite{Viyuela_etal2014a}.\label{fig:illo9a}]{%
		\includegraphics[width=0.35\textwidth]{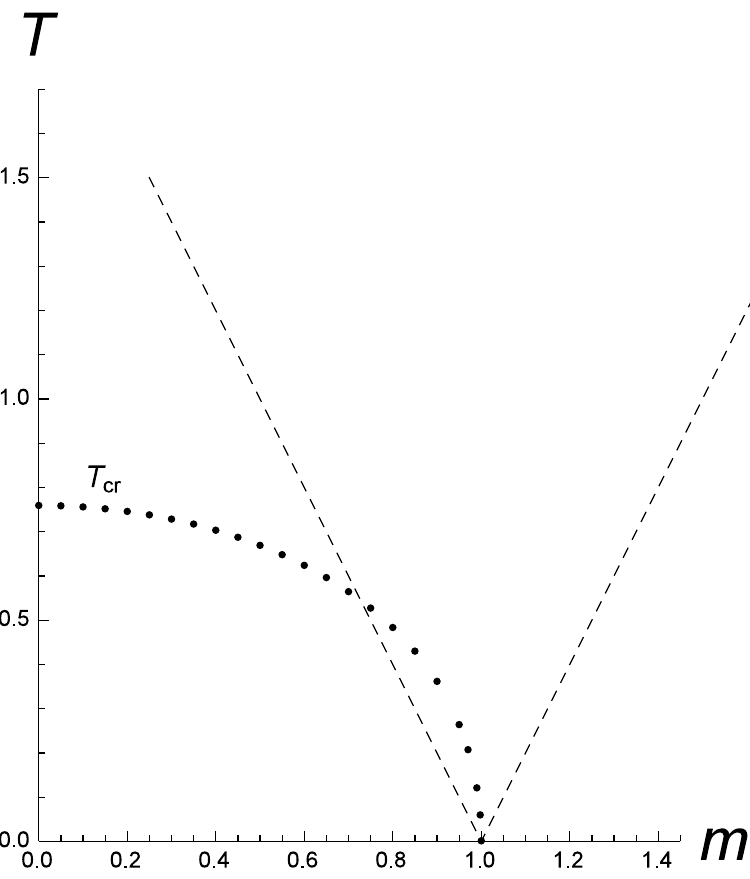}
	}
	~\hspace{20pt}
	\subfloat[Location of critical temperatures $T_{n_1,n_2}$ for $n_1 = 1$, 2 
	and 3 turns.\label{fig:illo9b}]{%
		\includegraphics[width=0.35\textwidth]{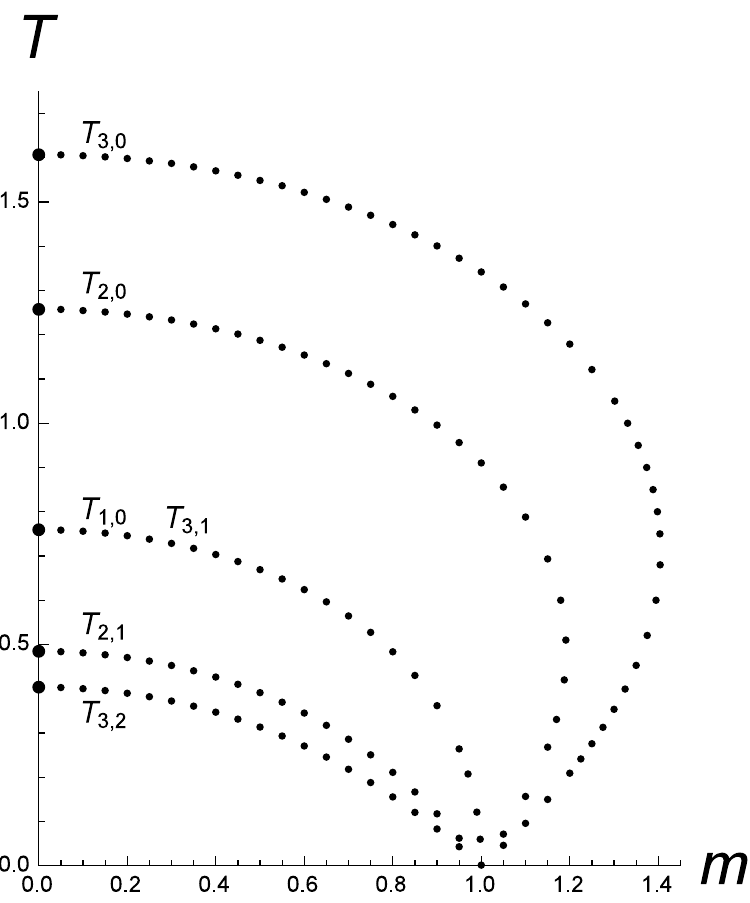}
	}
	\caption{\small Location of critical temperatures $T_{n_1,n_2}$ for $n_1 = 1$, 2 
		and 3 turns. Compare with Fig.~\ref{fig:illo4}, which shows the flat band case only. 
		The dashed line in panel (a) gives the size of the band gap (with Boltzmann's constant set to $1$).} 
	\label{fig:illo9}
\end{figure}

Going beyond the flat band case is a more involved story. We first ask for what 
value of $T$ we have a node occurring after exactly one turn. This means that we 
have to solve the equation 
\begin{equation} 
\cos{(A(T))} = 0,\qquad 
A(T) = \frac{1}{2}\int_0^{2\pi}\!\operatorname{d}\!k\,\dot{\phi}(1-x(k,T)). \label{Ab} 
\end{equation}
This is a straightforward Mathematica calculation, and the result is 
reported in Fig.~\ref{fig:illo9a}. 
The resulting curve is in fact a cross-section of 
any one out of the three two dimensional figures given by Viyuela \emph{et al.}~\cite{Viyuela_etal2014a} 
(their Figure 2, which is considerably more beautiful than ours). They interpret the resulting 
values of $T$---less than half of the bandgap at $m=0$---as an $m$-dependent critical 
temperature at which something dramatic 
happens to the Kitaev chain. We can, however, easily complicate matters by asking 
for the value of $T$ for which we have a node occurring after exactly two turns, 
or three turns and so on. Following this logic, we would then seem to be led to 
the conclusion that there is something dramatic happening to the Kitaev chain at 
finite temperature also for values of $m$ which exceed 1. See Fig.~\ref{fig:illo9b}. It is therefore necessary 
to reexamine the physical logic behind the Uhlmann phase, and we do so in Section \ref{discussion}.

\section{\large The interferometric geometric phase}\label{interGP}
In Ref:s.~\cite{Sjoqvist_etal2000,Tong_etal2004}, a different geometric phase for mixed quantum states was introduced in the context of interferometry.
Its general definition is fairly involved, but for mixed states having non-degenerate spectra, which is the only case considered here, the definition can be seen as a straightforward generalization of the Berry phase for pure states.

Consider a curve of density operators 
\begin{equation}
\rho(k)=\sum_ip_i(k)\ketbra{u_i(k)}{u_i(k)}
\end{equation}
such that for each $k$, the eigenvalues $p_i(k)$ are non-degenerate.  If the eigenstates develop in a parallel manner,
\begin{equation}\label{horizontal}
\braket{u_i(k)}{\dot{u}_i(k)}=0,
\end{equation} 
we define the interferometric geometric phase of $\rho(k)$ as
\begin{equation}\label{fas}
\gamma(k)=\arg\sum_i \sqrt{p_i(0)p_i(k)} \braket{u_i(0)}{u_i(k)}.
\end{equation}
As for Uhlmann's phase, the interferometric phase does not depend on the operator that determines the dynamics of the system. Rather, it is a quantity associated with $\rho(k)$ which only depends on how $\rho(k)$ is embedded in the space of density operators. However, this is true only if the eigenstates satisfy the parallelism condition \eqref{horizontal}. If this condition is not met, we need to phase-shift the eigenstates as follows:
\begin{align}
\ket{u_i(k)}\to \ket{u_i(k)}e^{-\int_0^k\operatorname{d}\!l \braket{u_i(l)}{\dot{u}_i(l)}}.
\end{align}
Clearly, the interferometric geometric phase reduces to the Berry phase if $\rho(k)$ is a curve of pure states. 

The Berry phase for a cyclically evolving pure state
equals half of the solid angle enclosed by the curve traced out by the state's Bloch vector on the Bloch sphere. Using this observation, it is fairly easy to calculate the interferometric phase for cyclically evolving qubits, provided they do not pass the maximally mixed state.
In fact, we believe that the parallelism condition in Eq.~\eqref{horizontal} cannot be extended to include all curves that pass through the maximally mixed state. However, this is still under investigation. 

Consider a closed curve of mixed qubits
\begin{equation}
\rho(k)=p_1(k)\ketbra{u_1(k)}{u_1(k)}+p_2(k)\ketbra{u_2(k)}{u_2(k)},\qquad (0\leq k\leq\kappa),
\end{equation}
and assume that the eigenvectors $\ket{u_i(k)}$ satisfy the parallelism condition \eqref{horizontal}.
Let $\vec{R}(k)$ be the Bloch vector of $\rho(k)$, and let $\vec{r}(k)=\vec{R}(k)/R(k)$, where $R(k)$ is the Euclidean length of $\vec{R}(k)$.
Then $\vec{r}(k)$ is a closed curve on the Bloch sphere and
$\braket{u_1(0)}{u_1(\kappa)}=e^{i\theta_1}$, where $\theta_1$ equals half of the solid angle enclosed by $\vec{r}(k)$. Furthermore, $\braket{u_2(0)}{u_2(\kappa)}=e^{i\theta_2}$, where $\theta_2$ is half of the solid angle enclosed by the curve $-\vec{r}(k)$. Accordingly, $e^{i\theta_2}=e^{-i\theta_1}$, and the interferometric geometric phase of $\rho(k)$ can be written as
\begin{equation}\label{fffas}
\gamma(\kappa)
= \arg(p_1(0)e^{i\theta_1}+p_2(0)e^{-i\theta_1})
= \arg(\cos\theta_1+iR(0)\sin\theta_1).
\end{equation}
We observe, in particular, that for the planar curves defined by \eqref{mixat} and indexed by $m$,
the interferometric phase is $0$ if the curve has an even winding number with respect to the maximally mixed state, and equals $\pi$ if the winding number is odd, see Fig.~\ref{fig:illo3}. 
When $m=1$, however, the interferometric  phase is undefined. Thus, it
signals a closing of the band gap for the Kitaev chain in the same way as the Berry phase, but 
at a finite temperature.

\begin{figure}[t]
\centering
\includegraphics[width=0.35\textwidth]{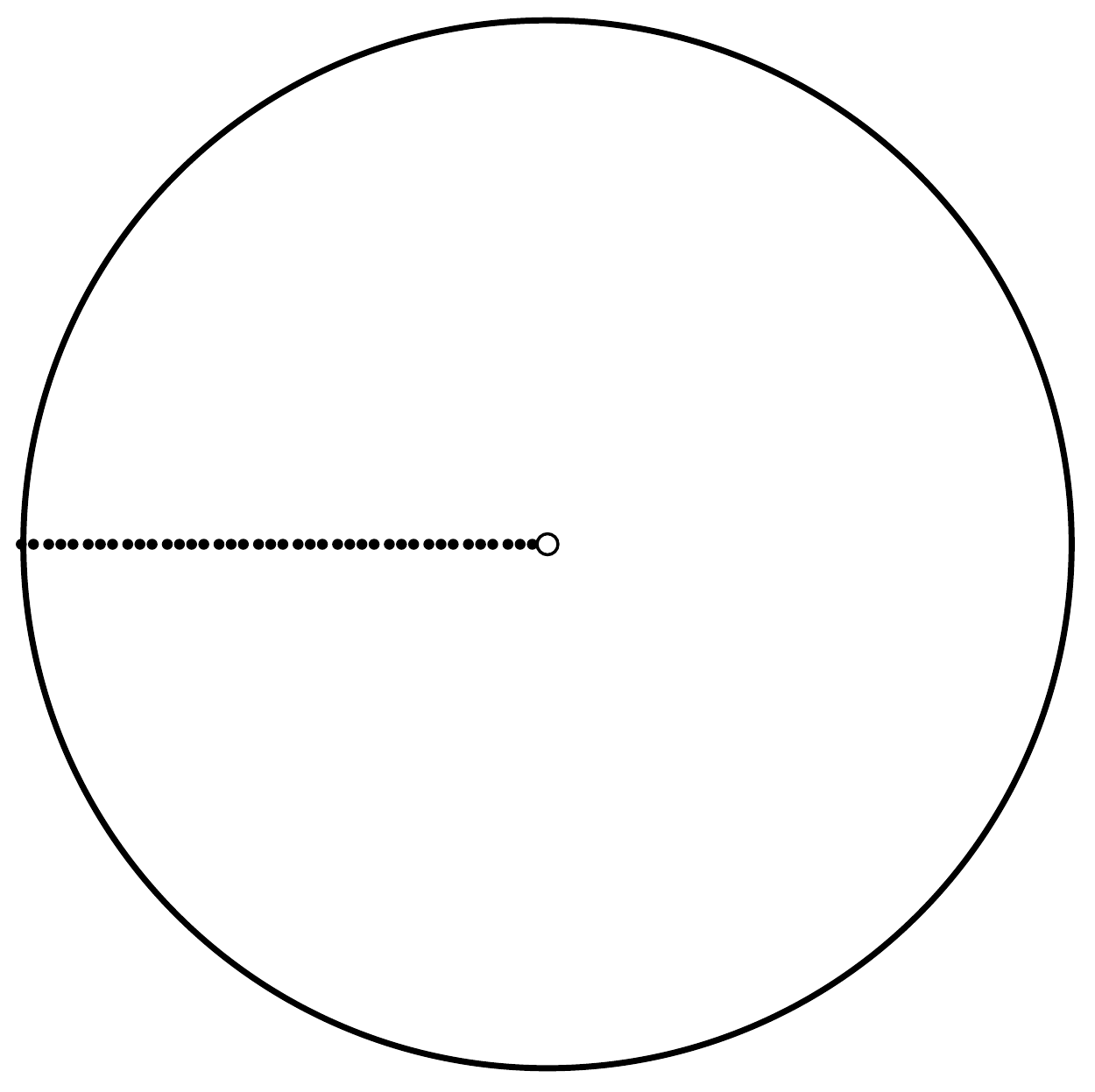}
\caption{\small The nodes of the interferometric phase are situated on the radial segment opposite to the initial state. No build-up of the phase occurs before and after the passage of a node. If the curve passes the line of nodes, the phase makes a sudden $\pi$-jump.}
\label{fig:ray}
\end{figure}

An analysis like the one conducted for Uhlmann's phase shows that the interferometric geometric phase has nodes along the radial segment in the Bloch ball opposite to the initial state. For the curves given by \eqref{mixat} this means that each point on the negative $x$-axis is a node, as shown in Fig.~\ref{fig:ray}. 
There is no phase build-up before, or after, the curve reaches a node. But at the node, the interferometric phase factor gets multiplied by $-1$. In this sense the behaviour of the interferometric phase closely mirrors that of the Berry phase.
 
\section{\large Discussion and conclusion}\label{discussion}
We have been concerned with the physical interpretation, in a
definite context, of two different geometric phases for mixed states---the Uhlmann and the interferometric phases.
Following the earlier studies \cite{Viyuela_etal2014a,Viyuela_etal2014c,Huang_etal2014,Budich_etal2015}, we have analyzed their behaviour to see
if they are indicators of topological phases of matter at finite temperature. As our test case we have used the Kitaev chain, a one-dimensional system of non-interacting electrons which gives rise to a curve of Gibbs states in an equatorial plane of the Bloch ball as we move along the Brillouin zone.

When $T=0$, and the Kitaev chain is in a pure state, the Berry phase is a topological invariant which takes different values around $m=1$ where the system undergoes a topological phase transition. For $m < 1$ the state is in a topologically non-trivial phase admitting localized zero-energy edge states, and the Berry phase factor is $-1$. For $m > 1$ there are no localized edge states and the Berry phase factor is $+1$. 
At finite temperature, on the other hand, the Kitaev chain is in a mixed state. 
Now, both of the mixed state geometric phase factors considered 
take the values $\pm 1$, but for the 
Uhlmann phase factor the behaviour is quite complex.  
We are not aware of a physical interpretation of the Uhlmann phase itself, but indeed the Uhlmann phase behaves in a way that may be of interest, regardless of the 
proposed interpretation of the states. Here it should be recalled that Uhlmann's theory was designed 
with very different ends in view. It can be argued that Uhlmann's $U(N)$ bundle is as 
natural a construction for mixed states as is the unit sphere in Hilbert space 
considered as a $U(1)$ bundle over the set of pure states. They both have major 
implications for the probabilistic structure 
of quantum mechanics \cite{Uhlmann_etal2009}. For example, Uhlmann's construction gives rise to a 
metric which is monotone under stochastic maps \cite{Uhlmann1992,Petz1996}, and plays an important role 
in statistical inference \cite{Braunstein_etal1994}. 
It must also be mentioned that the Uhlmann phase is measurable 
in experiments where the purified state appears as a bipartite entangled state 
including a system and a controlled ancilla \cite{Ericsson_etal2003,Aberg_etal2007,Zhu_etal2011}---but this is a very 
different situation from that of a system in contact with a thermal bath having a huge number of uncontrollable degrees of freedom.

What we found, when considering the concrete curves under study, 
was first of all an interesting, but in the context devastating, 'memory effect'. The phase changes along these 
curves occur only at certain nodes, and the locations of these nodes depend
on how many turns around the curve have been made already. 
Another observation concerns the limiting behaviour of the Uhlmann phase 
as the curve approaches the pure state boundary, which corresponds to letting the temperature approach zero. The 
Berry phase is recovered for closed curves \cite{Uhlmann1986}, but for open curves this 
is not the case. The argument is delicate because Uhlmann's bundle does not extend 
over the states of less than maximal rank (which play a special role in statistical inference). 

We now turn to the relevance of the Uhlmann phase for the condensed matter application 
we have in mind. Our first observation is that the Uhlmann phase is insensitive to the closure 
of the band gap. More precisely, if $T > 0$ and we let $m$ approach $1$ from below, we see that the Uhlmann phase changes its value before we reach $m=1$, see Fig.~\ref{fig:illo9a}.
When $m=1$, the band gap is closed, and the curve 
traversed by $\rho (k)$ passes through the maximally mixed state. However, this does not affect Uhlmann's geometric phase. In 
standard discussions of the topological origin of the edge states such behaviour of an invariant is 
explicitly forbidden (see, for instance, Ref.~\cite{Ryu_etal2002}). Furthermore, Fig.~\ref{fig:illo9b} shows that the Uhlmann phase behaves in a very non-trivial fashion for a range of temperatures also when $m>1$, in which case the chemical potential is in the non-topological range. 
We therefore agree with Viyuela {\it et al.} \cite{Viyuela_etal2014a} when they state that the 
Uhlmann phase ``does not determine the fate of the edge modes at finite temperature''. The Uhlmann phase was simply designed for a different purpose. 

In the spirit of Budich and Diehl \cite{Budich_etal2015}, we 
have also looked at a different geometric phase for mixed states, one that 
is sensitive to changes of the multiplicities in the spectra of the states. 
For the interferometric geometric phase, the phase factor is $-1$ for odd number of revolutions in the Brillouin zone and $+1$ for even numbers, no matter what the temperature is. This means that the interferometric phase does not detect any phase transition in temperature, but it detects the same phase transition as the Berry phase for pure states, i.e. at $m=1$. 
In fact, it extends the Berry phase to finite temperature in a way that captures the periodicity of the state of the Kitaev chain. For $\rho(k)$ given by Eq.~\eqref{mixat}, modeling pure states if $T=0$ and mixed states if $T>0$, induces a homomorphism between the fundamental groups of the Brillouin zone with end points identified and the equatorial plane punctured at the maximally mixed state. Both of these groups equal the group of integers, $\mathbb{Z}$, and the homomorphism is $n\to dn$ where $d$ is the degree of the Bloch vector. (Thus, $d=1$ if $m<1$ and $d=0$ if $m>1$.) We can extend this  with a second homomorphism $\mathbb{Z}\to\mathbb{Z}_2$ sending $n$ to $e^{in\pi}$. Here $\mathbb{Z}_2$ denotes the multiplicative group of the two elements $\pm 1$.
The composite homomorphism $n\to e^{idn\pi}$
is the topological invariant given by the Berry phase for $T=0$ and the interferometric phase for $T>0$. Observe that the non-periodicity, due to the memory effect mentioned above, excludes the possibility of extracting  a similar invariant from Uhlmann's geometric phase.

To summarize, the Uhlmann geometric phase is part of a general construction of great importance 
and import in quantum mechanics, but we see no reason for why it should be useful as an indicator 
of topological order in models resembling the Kitaev chain. The interferometric geometric phase on the other hand seems to be more appropriate as a topological order parameter at finite temperature, and we think it merits further investigation in the context of condensed matter.

\section*{\large Acknowledgments}
We thank Christian Sp\r{a}nsl\"{a}tt for patiently having answered all our questions about the physics of topological insulators and superconductors. E.S. acknowledges financial support from the Swedish Research Council.

\renewcommand*{\refname}{\vspace*{-1em}}

\end{document}